# A lightweight deep learning based cloud detection method for Sentinel-2A imagery fusing multi-scale spectral and spatial features

Jun Li, Zhaocong Wu, Zhongwen Hu, *Member, IEEE,* Canliang Jian, Shaojie Luo, Lichao Mou, Xiao Xiang Zhu, *Fellow, IEEE,* and Matthieu Molinier, *Member, IEEE*

*Abstract*—Clouds are a very important factor in the availability of optical remote sensing images. Recently, deep learning-based cloud detection methods have surpassed classical methods based on rules and physical models of clouds. However, most of these deep models are very large which limits their applicability and explainability, while other models do not make use of the full spectral information in multi-spectral images such as Sentinel-2. In this paper, we propose a lightweight network for cloud detection, fusing multi-scale spectral and spatial features (CD-FM3SF) and tailored for processing all spectral bands in Sentinel-2A images. The proposed method consists of an encoder and a decoder. In the encoder, three input branches are designed to handle spectral bands at their native resolution and extract multi-scale spectral features. Three novel components are designed: a mixed depth-wise separable convolution (MDSC) and a shared and dilated residual block (SDRB) to extract multi-scale spatial features, and a concatenation and sum (CS) operation to fuse multi-scale spectral and spatial features with little calculation and no additional parameters. The decoder of CD-FM3SF outputs three cloud masks at the same resolution as input bands to enhance the supervision information of small, middle and large clouds. To validate the performance of the proposed method, we manually labeled 36 Sentinel-2A scenes evenly distributed over mainland China. The experiment results demonstrate that CD-FM3SF outperforms traditional cloud detection methods and state-of-the-art deep learning-based methods in both accuracy and speed.

## I. Introduction

WHILE the spatial and spectral resolutions of optical remote sensing satellites increased in recent years, an inevitable problem has been influencing the application of optical remote sensing images: clouds. Clouds seriously influence the usability of optical remote sensing images [1]. The average cloud cover percentage of the Earth's surface is approximately 66% [2] which makes cloud detection and removal in optical remote sensing images a necessity.

Traditional cloud detection methods can be classified into three main types: threshold-based [3] or rule-based, object-based and multi-temporal based. These methods mainly use spectral information while ignoring spatial information that is very useful for distinguishing clouds from highlights such as snow/ice. Machine learning approaches such as SVM [4] and random forest [5] were adopted to improve cloud detection accuracy. Recently, convolutional neural networks (CNNs) in deep learning have been widely used for feature extraction especially spatial features in image processing. CNN has also been introduced into cloud detection in remote sensing images and achieved high performance [6]. However, deep learning-based methods usually have tens of millions of parameters and require high performance hardware which limits their application. Although dilated convolution has been widely used for reducing parameters and obtaining large receptive field, it causes grid effects.

The performances of cloud detection methods are usually validated on Landsat and Sentinel-2 imagery since they are widely used for monitoring the land resources and environmental changes [7], [8]. Fmask [9] is a rule-based cloud detection method that was originally developed for Landsat imagery, then adapted to Sentinel-2 imagery [10]. In Fmask, several rules designed by experience and knowledge of spectral characteristics are used to distinguish clouds from some land cover types such as water and snow. The results of these rules are combined to produce a final cloud mask, then considering

This work was supported in part by the National Natural Science Foundation of China (NSFC) under Grant 41871227, in part by the Natural Science Foundation of Guangdong under Grant 2020A1515010678, in part by the State Grid Technology Project under Grant 5700-202019162A-0-0-00, in part by the Academy of Finland through the Finnish Flagship Programme FCAI: Finnish Center for Artificial Intelligence under Grant 320183 and in part by VTT through the project S5G4I/EdgeAI. (*Corresponding author: Zhaocong Wu.*)

J. Li and Z. Wu are with the School of Remote Sensing and Information Engineering, Wuhan University, Wuhan 430079 China e-mail: zcwoo@whu.edu.cn; jun__li@whu.edu.cn).

Z. Hu is with the MNR Key Laboratory for Geo-Environmental Monitoring of Great Bay Area, Shenzhen University, Shenzhen 518060 China (e-mail: zwhoo@szu.edu.cn).

C. Jian is with The Administration Department, Fujian Tendering Purchasing Group Co., Ltd, Fuzhou 350001 China (e-mail: jiancanliang@163.net).

S. Luo is with the Hangzhou Power Supply Company, State Grid Zhejiang Power Supply Company, Hangzhou 310000 China (e-mail: shaojieluo@sina.com).

L. Mou and X. X. Zhu are with the Remote Sensing Technology Institute, German Aerospace Center, 82234 Wessling, Germany and also with the Data Science in Earth Observation (SiPEO, former: Signal Processing in Earth Observation), Technical University of Munich, 80333 Munich Germany (e-mail: xiaoxiang.zhu@dlr.de; lichaomou@dlr.de).

M. Molinier is with VTT Technical Research Centre of Finland Ltd, Espoo 02044 Finland (e-mail: matthieu.molinier@vtt.fi).

the sizes and heights of clouds, sun angle and auxiliary geometry data, cloud shadows are inferred from cloud masks and detected. In order to separate clouds from highlights in Sentinel-2 images, NIR parallax which uses the angular characteristics of bands 7, 8 and 8A that are observed from different view angles was integrated into Fmask [11]. Recently, Fmask 4.0 was developed and achieved 94.3% cloud detection accuracy in Sentinel-2 imagery [12]. It uses Global Surface Water Occurrence (GSWO) to better separate land and water, and a global Digital Elevation Model (DEM) to reduce the influence of elevation variations on cirrus bands.

However, rule-based based methods are limited by the spectral bands and do not perform well on bright surfaces. Multi-temporal based methods using changed information in image time-series to detect clouds may improve cloud detection accuracy on bright surfaces even with limited bands [13]–[16]. The main limitation of multi-temporal based methods is that the processing time of image time-series is significantly more than that of a single image. It should also be noted that many multi-temporal based methods require a clear image within the specific time period, which is not always available within a short time in the most frequently cloudy areas such as tropical or boreal zones.

In recent years, machine learning-based cloud detection methods have been widely investigated in remote sensing field [17]–[19]. In [5], the spectral indices of different land cover types were first calculated and put into a random forest to produce preliminary cloud masks, then fed into a Super-pixels Extracted via Energy-Driven Sampling (SEEDS) segmentation algorithm to get the final masks. Cilli *et al*. [20] compared six classically used methods: Support Vector Machines (SVM), Random Forests (RF), Multilayered Perceptron (MLP), Sen2Cor [21], Fmask and MAJA [16] for cloud detection in Sentinel-2 images and concluded that SVM was the best option among these machine learning-based and model-based methods.

As a branch of machine learning, deep learning (DL) is developing very fast and has been widely used in many applications of remote sensing, such as land use mapping [22] hyperspectral classification [23] cloud removal [24] [25] and missing data reconstruction [26]. Due to its good performance on image classification, deep learning has also been adopted to improve the accuracy of cloud detection in remote sensing images [27]–[31]. Shao *et al*. [32] stacked 10 Landsat-8 multi-spectral bands after resampling and designed four parallel average pooling filters to obtain multiscale features of clouds. The RGB thumbnails of ZY-3 images were first used for cloud detection in [33] and RGB bands in Landsat-8 Biome [34] and in GF1_WHU [35] datasets were also used for the validation of the method. In [36], RGB bands in GF-1 images were converted into 8-bit and down-sampled to 512×512 to train their model, and the same pre-processing step was also applied to Landsat-8 Biome and GF1_WHU datasets for testing. A U-net based method was proposed in [37], in which cloud masks produced by Fmask were first used for model training, and the well-trained model performed better than Fmask. Li *et al*. [38] proposed a multi-scale convolutional feature fusion (MSCFF) network which up-samples the multiple high level features then fuses them and low level features to produce final masks. This method was quantitatively validated on satellite images of different spatial resolutions: Landsat-7 Irish [39], Landsat-8 Biome, GF1_WHU and HRC_WHU [38] datasets.

Recently, semi and weakly supervised methods have been widely used to reduce the manual labor in labelling dataset [40]–[42], such as replacing pixel-level labels with image-level labels during model training [43]–[45]. Weakly supervised learning strategy is also used for cloud detection in remote sensing images [46], since it is very convenient to produce training labels in a fast and reliable way. In SAGAN [44], self-attention mechanism and generative adversarial network (GAN) [47] were combined to detect clouds in Sentinel-2 images. SAGAN takes images and corresponding image-level binary labels as inputs (1 for cloud image, 0 for cloud-free images - which are easier to collect than pixel-level labels) and produces cloud masks at pixel-level by extracting the difference between cloud and cloud-free images. A mixed energy separation based method was proposed in [48], in which cloud layers were first generated by inputting cloud and cloud-free images into a cloud generator. In the next step, the cloud layers were fused with the cloud-free image to produce a fake cloud image, which was then put into a cloud detection network. Finally the consistency between the cloud layers and cloud detection results were used to supervise the whole model.

Many deep learning-based methods have been proposed for cloud detection in remote sensing images and achieved the state-of-the-art performance. However, most of these models tend to be very large, with millions of parameters, which hinders their applicability and explainability. Furthermore, it is still difficult to distinguish clouds from highlights such as buildings, barren and snow/ice with only visible and near infrared (VNIR) bands. In recent years, many high-resolution satellites equipped with multi-spectral sensors have been widely used for land resource monitoring, such as Sentinel-2A/B, CBERS-04, ZY-1 02D and HJ-1B. These satellites include short wave infrared (SWIR) bands at lower spatial resolutions than VNIR bands. Although SWIR bands include useful spectral information to distinguish clouds from highlights, most deep learning-based cloud detection methods only handle high resolution VNIR bands and ignore other low resolution bands, such as SWIR bands.

Unlike previous works, we processed all Sentinel-2A spectral bands simultaneously in the proposed CD-FM3SF method. Quantitatively and qualitative comparisons were run with classical cloud detection methods, Fmask 4.0 and Sen2cor, as well as two deep learning-based methods MSCFF [38] and RS-Net [37] sharing similar basic architecture or components with CD-FM3SF. An ablation experiment was conducted to evaluate the performance of CD-FM3SF when only VNIR bands were used as inputs, such as in baseline deep learning methods. Comparisons on WHUS2-CD dataset demonstrate the superiority of the proposed CD-FM3SF method, both in terms of accuracy and inference speed, whether all spectral bands or only VNIR bands were used as inputs. In this work, we propose a lightweight deep learning-based network called CD-FM3SF (Cloud Detection method Fusing Multi-Scale Spectral and Spatial Features) to detect clouds in Sentinel-2A (S2) images. The model is designed to make full use of the multi-spectral information in Sentinel-2A images to improve cloud detection accuracy. With all bands as inputs, the model can produce more accurate cloud masks even if most bands have lower spatial resolution than VNIR bands. The main contributions of this work are as follows:

1) A lightweight deep learning-based model based on the residual and encoder-decoder architectures is designed for processing all bands of Sentinel-2A images. Three learnable CNN layers are used to handle bands at different resolutions. By integrating the SWIR spectral features, clouds can be easily distinguished from snow/ice.

2) Three novelty elements are designed to reduce the number of parameters and fuse multi-scale spectral and spatial features: a mixed depth-wise separable convolution (MDSC), a shared and dilated residual block (SDRB) and a Concatenate and Sum (CS) operation.

3) A new Sentinel-2 cloud detection dataset (WHUS2-CD) containing 32 Sentinel-2A images is introduced, covering all seasons and various land cover types over mainland China. Reference cloud masks in WHUS2-CD are labeled at the highest available spatial resolution, 10 m, higher than exiting S2 cloud datasets. The dataset is available online in open access (https://github.com/Neooolee/WHUS2-CD).

The rest of this paper is organized as follows. Section II presents the experiment data. Section III introduces the proposed CD-FM3SF method. The experimental results are shown in Section IV and Discussions is made in Section V. Conclusions are presented in Section VI.

## II. DATA

### A. Sentinel-2A Imagery and Spectral Characteristics for Cloud Detection

Sentinel-2 is a European wide-swath, high-resolution, multi-spectral imaging mission, that includes two satellites operating in opposite sides of the orbit to provide a revisiting time of up to 5 days. Like Landsat imagery, Sentinel-2 images have been widely used for land cover mapping, vegetation monitoring and disaster assessment due to its convenient open access policy and high-quality images. Equipped with a multi-spectral imaging (MSI) system covering 13 spectral bands in the 0.443 μm to 2.190 μm range (Table I), Sentinel-2 is the only high-resolution satellite that includes 3 bands in vegetation red edge which are very useful for vegetation monitoring. The resolutions of Sentinel-2A bands are 10 m, 20 m and 60 m, with higher resolution bands used for land cover mapping and environment monitoring, and the three 60 m resolution bands used to detect atmospheric vapour [49].

Fig. 1 shows the TOA reflectance of five typical land cover types and clouds on each S2 band. The curves were obtained by randomly sampling 8 pixels for each land cover type and averaging TOA reflectance band by band. It can be seen that vegetation, water, urban and barren classes have quite different TOA values from clouds on most bands. Thus, it is easy to distinguish clouds from these land cover types. Snow/ice

### TABLE I
SENTINEL-2A SENSOR BANDS. ALL BANDS WERE USED IN OUR STUDY.

| Band no. | Band name | Central Wavelength (μm) | Bandwidth (nm) | Spatial resolution (m) |
|---|---|---|---|---|
| Band 1 | Coastal aerosol | 0.443 | 27 | 60 |
| Band 2 | Blue | 0.490 | 98 | 10 |
| Band 3 | Green | 0.560 | 45 | 10 |
| Band 4 | Red | 0.665 | 38 | 10 |
| Band 5 | Vegetation Red Edge | 0.705 | 19 | 20 |
| Band 6 | Vegetation Red Edge | 0.740 | 18 | 20 |
| Band 7 | Vegetation Red Edge | 0.783 | 28 | 20 |
| Band 8 | NIR | 0.842 | 145 | 10 |
| Band 8A | Vegetation Red Edge | 0.865 | 33 | 20 |
| Band 9 | Water Vapour | 0.945 | 26 | 60 |
| Band 10 | SWIR-Cirrus | 1.375 | 75 | 60 |
| Band 11 | SWIR | 1.610 | 143 | 20 |
| Band 12 | SWIR | 2.190 | 242 | 20 |

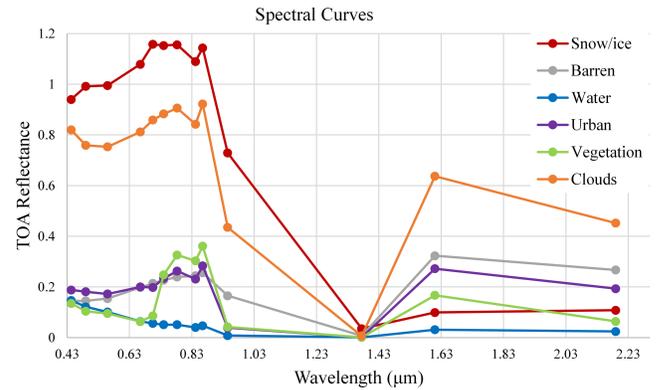

Fig. 1. TOA reflectance of five typical land cover types and clouds.

reflects most incident radiation in bands 1 to 10 (from 0.443 μm to 1.375 μm), while absorbing most incident radiation in bands 11/12 (SWIR). Therefore, the TOA reflectance of snow/ice is high in visible bands but drops sharply in SWIR. The spectral reflection of snow is very similar to that of clouds in bands 1 to 10, but very different in bands 11/12, which indicates that spectral information in SWIR bands can be used for distinguishing snow from clouds. Based on this, we proposed a new deep learning-based model CD-FM3SF which takes all bands as input when detecting clouds in Sentinel-2A images.

### B. WHUS2-CD and WHUS2-CD+ Cloud Detection Datasets

Three Sentinel-2 cloud detection datasets are publicly available [50]–[52]. Although these datasets are widely distributed, the reference cloud masks are only at 20 m [50], [52] or 60 m [51] resolution, and do not benefit from the 10 m VNIR bands in Sentinel-2 image. Furthermore, the original image source files used in creating the cloud masks are not included in the datasets, which makes them less practical to use as benchmark datasets. We propose to address these limitations by providing an open Sentinel-2 cloud mask dataset at the highest resolution (10m) that includes the input S2 image files. In this work, mainland China was selected as the experiment

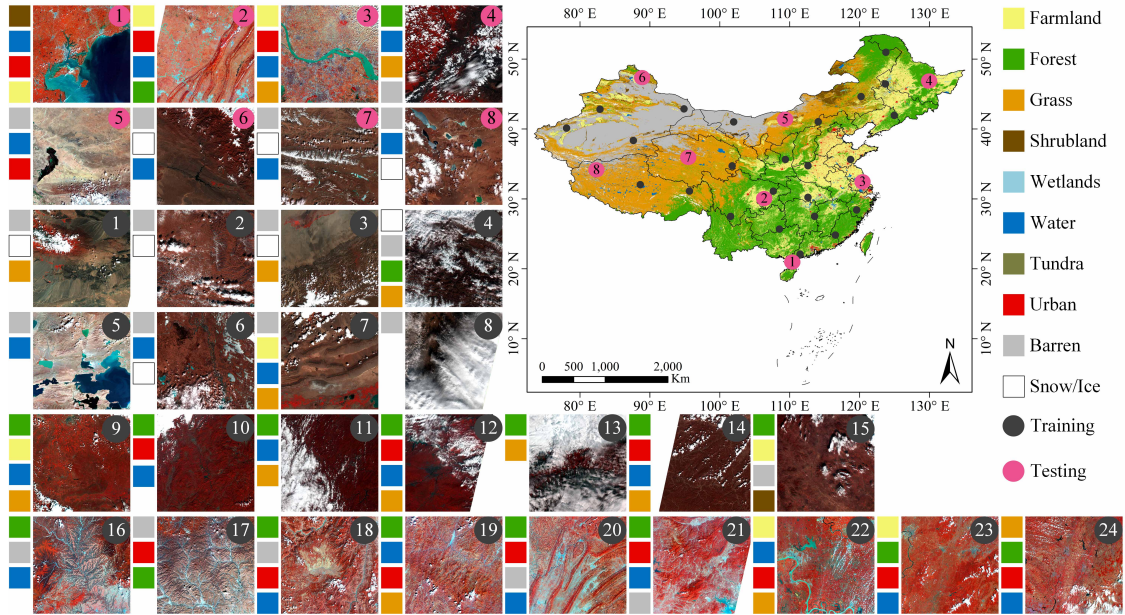

Fig. 2. WHUS2-CD dataset: 8 testing images (first two rows) and 24 training images. Colour labels on the left of each image show their main land cover. The number in each region corresponds to the image index in Table II. Land cover types from ESA-CCI-LC [53].

TABLE II
DETAILS OF TRAINING AND TESTING IMAGES IN WHUS2-CD DATASET.

| | No | Product ID | Main land cover types | Acquisition date |
|---|---|---|---|---|
| Training | 1 | S2A_MSIL1C_20190714T043711_N0208_R033_T46TFN_20190714T073938 | Barren/Snow/Ice/grass | 2019-07-14 |
| | 2 | S2A_MSIL1C_20191219T040151_N0208_R004_T47SQU_20191219T055033 | Barren/Snow/Ice | 2019-12-19 |
| | 3 | S2A_MSIL1C_20190630T045701_N0207_R119_T45SWC_20190630T080543 | Snow/Ice/Barren/Grass | 2019-06-30 |
| | 4 | S2A_MSIL1C_20191215T042551_N0208_R090_T46RGV_20191215T065406 | Snow/Ice/Barren/Forest/Grass | 2019-12-15 |
| | 5 | S2A_MSIL1C_20180930T044701_N0206_R076_T45SXR_20180930T074413 | Barren/Water | 2018-09-30 |
| | 6 | S2A_MSIL1C_20200317T024541_N0209_R132_T51TWM_20200317T053350 | Barren/Water/Snow/Ice | 2020-03-17 |
| | 7 | S2A_MSIL1C_20180816T053641_N0206_R005_T44TKK_20180816T093424 | Barren/Farmland/Water/Grass | 2018-08-16 |
| | 8 | S2A_MSIL1C_20191023T040821_N0208_R047_T47TQF_20191023T074550 | Barren | 2019-10-23 |
| | 9 | S2A_MSIL1C_20180824T031541_N0206_R118_T50TKL_20180824T061636 | Forest/Farmland/Water/Grass | 2018-08-24 |
| | 10 | S2A_MSIL1C_20191118T025011_N0208_R132_T50RMN_20191118T071843 | Forest/Urban/Water | 2019-11-18 |
| | 11 | S2A_MSIL1C_20190916T023551_N0208_R089_T50RQS_20190916T042547 | Forest/Water/Grass | 2019-09-16 |
| | 12 | S2A_MSIL1C_20190819T031541_N0208_R118_T49SFU_20190819T065332 | Forest/Urban/Water/Grass | 2019-08-19 |
| | 13 | S2A_MSIL1C_20190815T051651_N0208_R062_T44TPN_20190815T090034 | Forest/Grass | 2019-08-15 |
| | 14 | S2A_MSIL1C_20200410T022551_N0209_R046_T51TXG_20200410T042047 | Forest/Urban/Water/Grass | 2020-04-10 |
| | 15 | S2A_MSIL1C_20191002T025551_N0208_R032_T50TQQ_20191002T054113 | Forest/Farmland/Barren/Shrubland | 2019-10-02 |
| | 16 | S2A_MSIL1C_20180429T032541_N0206_R018_T49SCV_20180429T062304 | Forest/Barren/Water | 2018-04-29 |
| | 17 | S2A_MSIL1C_20200506T024551_N0209_R132_T51UWS_20200506T043639 | Barren/Urban/Forest | 2020-05-06 |
| | 18 | S2A_MSIL1C_20200325T034531_N0209_R104_T47RQL_20200325T065315 | Forest/Barren/Urban/Water | 2020-03-25 |
| | 19 | S2A_MSIL1C_20190928T031541_N0208_R118_T49RBJ_20190928T061248 | Forest/Water/Urban/Grass | 2019-09-28 |
| | 20 | S2A_MSIL1C_20180827T032541_N0206_R018_T48RYV_20180827T062627 | Forest/Urban/Barren/Water | 2018-08-27 |
| | 21 | S2A_MSIL1C_20200222T030731_N0209_R075_T49QEE_20200222T060244 | Forest/Urban/Water/Barren | 2020-02-22 |
| | 22 | S2A_MSIL1C_20180722T030541_N0206_R075_T49RFP_20180722T060550 | Farmland/Water/Urban/Grass | 2018-07-22 |
| | 23 | S2A_MSIL1C_20180729T025551_N0208_R032_T49RGL_20180729T055945 | Farmland/Forest/Urban/Water | 2018-07-29 |
| | 24 | S2A_MSIL1C_20200506T024551_N0209_R132_T50SPE_20200506T052918 | Grass/Forest/Urban/Water | 2020-05-06 |
| Testing | 1 | S2A_MSIL1C_20180930T030541_N0206_R075_T49QDD_20180930T060706 | Shrubland/Water/Urban/Barren | 2018-09-30 |
| | 2 | S2A_MSIL1C_20191105T023901_N0208_R089_T51STR_20191105T054744 | Farmland/Urban/Water/Forest | 2019-11-05 |
| | 3 | S2A_MSIL1C_20190812T032541_N0208_R018_T48RXU_20190812T070322 | Farmland/Urban/Water/Grass | 2019-08-12 |
| | 4 | S2A_MSIL1C_20190602T021611_N0207_R003_T52TES_20190602T042019 | Forest/Water/Grass/Barren | 2019-06-02 |
| | 5 | S2A_MSIL1C_20190328T031731_N0207_R061_T49TCF_20190328T071457 | Barren/Water | 2019-03-28 |
| | 6 | S2A_MSIL1C_20191001T050701_N0208_R019_T45TXN_20191002T142939 | Barren/Snow/Ice/Water | 2019-10-02 |
| | 7 | S2A_MSIL1C_20200416T042001_N0209_R133_T46SFE_20200416T074050 | Barren/Snow/Ice/Water | 2020-04-16 |
| | 8 | S2A_MSIL1C_20200528T050701_N0209_R019_T44SPC_20200528T082127 | Barren/Water/Snow/Ice | 2020-05-28 |

area and a cloud detection validation dataset called WHUS2-CD is presented. The main land cover types over the globe were classified into 10 level-1 classes: farmland, forest, grass, shrubland, wetlands, water, tundra, urban, barren and snow/ice in [54]. As shown in Fig. 2, we made sure the study regions were evenly distributed over mainland China while taking land cover types into consideration. We also set the time span of the experiment data from April 2018 to May 2020 in order to cover all seasons. With these conditions, 32 Sentinel-2A images covering about 380000 km² of Earth surface were finally

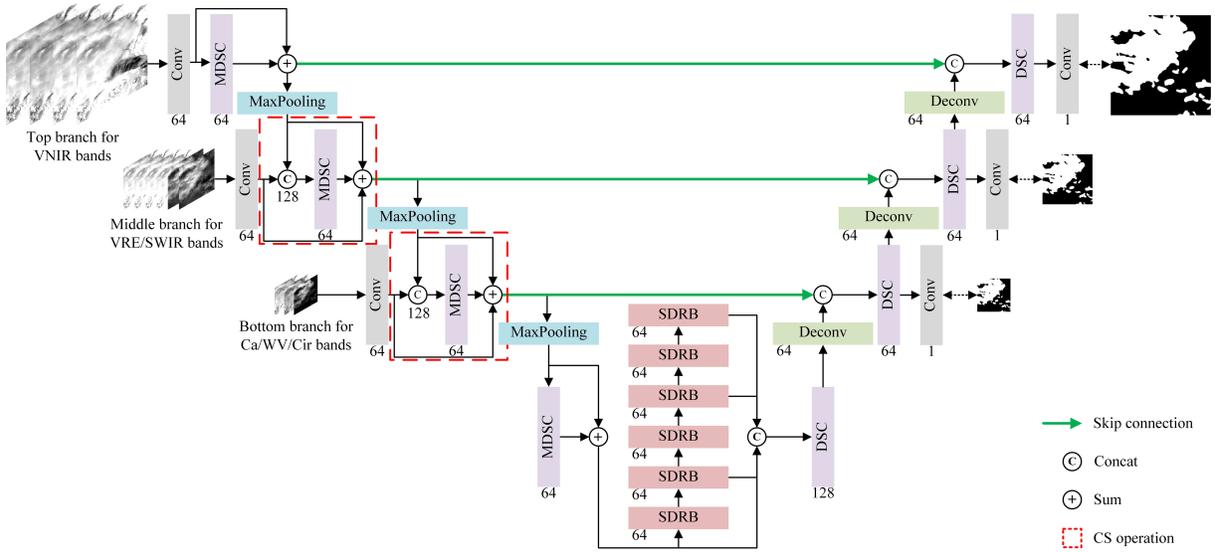

Fig. 3. The architecture of CD-FM3SF. There are three branches used to process VNIR, VRE/SWIR, and CA/WV/Cir bands respectively. The number under each layer is the number of its output feature maps.

selected as the experiment data in WHUS2-CD (Table II).

In order to better test the performance of CD-FM3SF and compared methods, the test dataset is chosen evenly distributed over China mainland, including various land cover types (vegetation, water, urban, barren and snow/ice) and all seasons. Since in most practical applications, images with more than 20% cloud cover are not used, cloud detection accuracy on images with less than 20% cloud cover is much more relevant. Therefore, we selected more images with less than 20% cloud cover in the test dataset. In order to evaluate CD-FM3SF accuracy also on images with high cloud cover, we selected one image with about 65% cloud cover. Since the biggest challenge of cloud detection is distinguishing clouds from highlights, especially barren and snow/ice, we selected 1 barren image and 3 snow/ice images. Cloud detection on urban areas is more difficult than on vegetation, so 3 images containing urban area were included. Finally, 8 images from the 32 Sentinel-2A images satisfying these the conditions were selected as the test images, covering about 86000 km² of Earth surface. The other 24 images were used for training deep learning-based models. The distributions and main land cover types of the training and testing datasets is shown on Fig. 2. Since distinguishing clouds from snow/ice is the biggest challenge in cloud detection, 4 additional snow/ice images were selected for further validation of cloud detection methods and added to WHUS2-CD to create WHUS2-CD+ dataset.

The reference cloud masks in WHUS2-CD+ were manually labeled at 10 m spatial resolution, the highest available from Sentinel-2A images. Therefore, the performance of cloud detection methods can be tested with greater accuracy than with the two Sentinel-2 validation cloud masks mentioned in the related works (20 m). The false color composited images of bands 8/3/2 were mainly used to delineate cloud masks. In images containing snow, the false color composited image of bands 8A/11/12 was used to distinguish clouds from snow. All false color composited images were stretched by a 2% liner stretch for better visualization.

Adobe Photoshop software was used to produce cloud masks. All cloud boundaries were manually labeled on a transparent layer with lasso and magic wand tools. Then the selected cloud pixel values were assigned to 255 value. A background layer with all pixel values as 128 was created and then composited with the cloud layer to get the final cloud mask. As in [35], thin clouds whose boundaries were visually identifiable and below which background could not be seen clearly were labeled as opaque clouds.

### C. Details of training Data for different methods

The Sentinel-2A level 1C top-of-atmosphere reflectance (TOA) product was used as the experimental data for all methods. All baseline deep learning-based methods only used VNIR bands as inputs due to the limitations of their architectures. Inputs for Fmask are bands 2/3/4/7/8/8A/10/11/12 and for Sen2Cor bands 1/2/3/4/5/8/8A/10/11/12, whereas the proposed CD-FM3SF takes all Sentinel-2A bands as inputs. For deep learning-based methods, all input images were normalized between 0 and 1 by dividing TOA reflectance by 10000.

In addition, in order to analyze the influence of atmospheric vapour bands on the performance of CD-FM3SF, we conducted ablation experiments without CA/WV/Cir bands as inputs (CD-FM3SF-10). Because many optical satellite sensors include VNIR bands, an experiment with only VNIR bands as inputs (CD-FM3SF-4) was conducted to demonstrate the potential of CD-FM3SF for cloud detection in other optical satellite images. This also allows a fair comparison to original baseline deep learning methods with only VNIR bands as inputs.

For a more fair comparison, we also rescaled 20 m and 60 m bands into 10 m by bi-linear interpolation and used all 13 bands as input channels of first layer to 13 in deep learning-based baseline methods. The original deep learning-based baseline methods with only VNIR bands as inputs are called MSCFF and RS-Net. The modified MSCFF and RS-Net for processing all bands are called MSCFF-13 and RS-Net-13.

Due to memory limitations of the hardware for deep learning-based methods, we cropped the 24 training images into small patches. Slide windows of different sizes were used, with half overlapping. Since the spatial resolution of VNIR, VRE/SWIR and Ca/WV/Cir bands are 10 m, 20 m and 60 m, the corresponding slide window sizes were set to 384×384 , 192×192 and 64×64 pixels, respectively, corresponding to 192×192 , 96×96 and 32×32 overlapping pixels. All slide windows had the same footprint so that the area extents of patches from different bands matched with each other. Finally, 73429 grouped multi-resolution image patches were produced after deleting patches with no data.

CD-FM3SF uses three scale cloud masks for supervision, that are prepared as follows. We applied a 2×2 and a 6×6 slide windows on 10 m resolution reference masks, then assigned the majority class in the window. In this way, reference masks at 20 m and 60 m resolution were produced. Finally, we cropped the 10 m, 20 m and 60 m cloud masks with the same strategy as for cropping image patches.

We used data augmentation on input patches, by randomly flipping horizontally and vertically, and rotating at 90°, 180° and 270° angles. This effectively increased the training data size by a factor of 5, since transformed images are treated as new data by deep learning-based models. Thus, 367145 grouped training image patches were produced for training of deep learning-based methods.

## III. METHODOLOGY

### A. Overview of CD-FM3SF

Fig. 3 shows an overview of CD-FM3SF, with three inputs from multi-spectral image stacks at different resolutions: VNIR composited with bands 2/3/4/8 (Visible and Near-Infrared), VRE/SWIR composited with bands 5/6/7/8A/11/12 (Vegetation Red Edge and Short-Wave Infrared) and Ca/WV/Cir composited with bands 1/9/10 (Coastal Aerosol, Water Vapour and Cirrus). These band stacks are inputs to top, middle and bottom branches respectively in training and testing. The features of these multi-spectral images are fused together after being processed by the corresponding input branches. CD-FM3SF builds upon the residual architecture [55], which has been proved very beneficial for model training. It should be noted that a new residual architecture called CS (concatenation and sum) was designed in CD-FM3SF to better fuse multi-scale spatial and spectral features. Furthermore, the parameters in CD-FM3SF are greatly reduced by replacing normal convolutions with depth-wise separable convolutions (DSC) [56]. In addition, a novel mixed depth-wise separable convolution (MDSC) combining group convolution [57] and DSC was designed to fuse multi-scale spatial features from multi-spectral bands in the encoder. A residual block that combines shared convolutions and dilated convolutions (SDRB) was designed to solve the gridding effect cause by dilated convolutions. These layers are used to extract and fuse multi-scale spatial and spectral features which are very effective for the improvement of CD-FM3SF on cloud detection. The following subsections give more details about the components of CD-FM3SF.

CD-FM3SF consists of many basic layers: convolution layer, deconvolution layer, depth-wise separable convolution layer (DSC), mixed depth-wise separable convolution layer (MDSC), shared convolution layer, dilated convolution layer, and shared convolution and dilated convolution residual block (SDRB). Out of those components, MDSC and SDRB are novel and original contributions of CD-FM3SF.

### B. Normal Layers

*1) The convolution layer* (Conv) contains multiple convolution kernels and is used to extract features from input data. Each element that constitutes the convolution kernel corresponds to a weight coefficient and a bias. Each neuron in the convolution layer is connected with multiple neurons in the adjacent region from the previous layer. Noting h the height, w the width, b the batch size and m the number of input channels, the input shape is b×h×w×m. If the number of output channels is n, the number of filters is m×$n$. Assuming that the kernel size of each filter is k×k and a bias is used in each filter, the number of parameters of a normal convolution layer is:

$$N_{Conv}=m×k×k×n+m×n \qquad (1)$$

In CD-FM3SF, every convolution layer is followed by rectified linear unit (ReLU) activation function.

*2) The deconvolution layer* (Deconv) is used to up-sample the input feature maps, by interpolating zero values between the elements of the input matrix according to the input stride, then applying a normal convolution operation with stride = 1. With the same input shape b×h×w×m and assumptions for kernel size, bias and output channels as for normal convolution layers, the number of parameters of a normal Deconv layer is:

$$N_{Deconv}=m×k×k×n+m×n \qquad (2)$$

We set the kernel size of all Deconv layer in CD-MF3SF to 4×4 and the strides were set to 2, 3 and 2 from bottom to top.

*3) The pooling layer* is used to down-sample the input feature maps to a certain size without any parameters. The two most used pooling strategies are max-pooling and average pooling. Assuming that the window size of pooling is k×k, max-pooling and average pooling will output the maximum and average values of the elements in k×k window, respectively. As in many deep learning-based methods [57]–[59], three max-pooling are adopted in CD-FM3SF to preserve texture features when down-sampling the feature maps (Fig. 3). The window sizes of max-pooling layers are set to 2×2, 3×3 and 2×2, respectively. The strides of max-pooling layers are set 2, 3 and 2, respectively.

*4) The depth-wise separable convolution layer* (DSC) is the depth-by-depth convolution, or the channel-by-channel convolution [56]. As shown in Fig. 4 (a), the number of filters in DSC corresponds to the number of input channels. Assuming m input and n output channels, first m filters with kernel size k×k are applied to m input feature maps to produce m corresponding feature maps. Then, m×n filters with kernel size 1×1 are applied to the m intermediary feature maps to produce

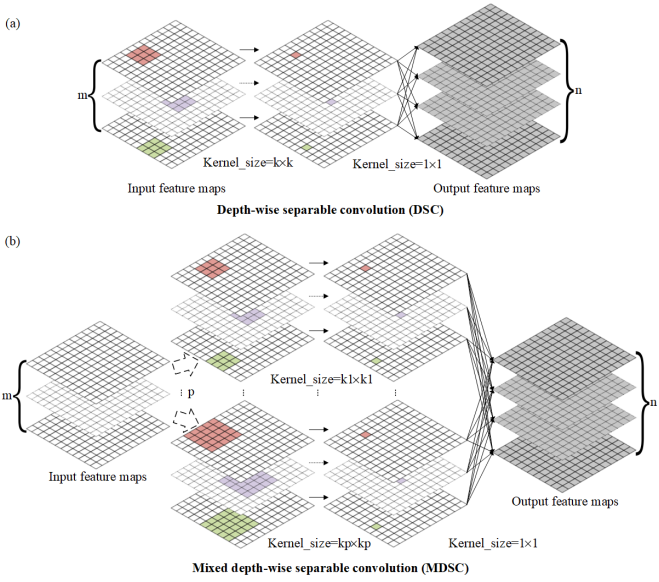

Fig. 4. The details of depth-wise separable convolution layer (DSC) and mixed depth-wise separable convolution layer (MDSC). The feature maps in each dotted box are concatenated at feature channels.

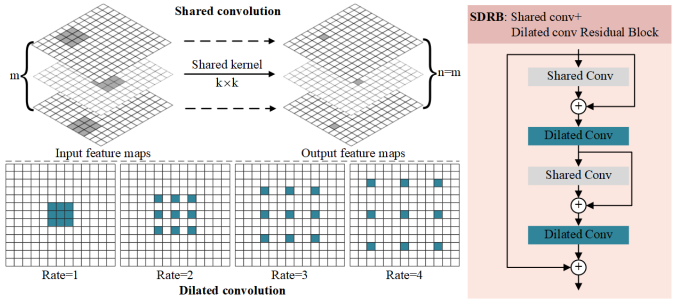

Fig. 5. The details of shared and dilated convolution residual block layer (SDRB) and mixed depth-wise convolution layer (MDSC). The feature maps in each dotted box are concatenated at feature channels.

n output feature maps. The total number of parameters of a DSC layer is:

$$N_{DSC} = m \times k \times k + m \times n \qquad (3)$$

In CD-FM3SF, k is set to 3.

*5) The shared convolution layer* is used to extract spatial information in different feature maps using the same filter. As shown on the top left of Fig. 5, there is only 1 filter with kernel size k×k in *shared convolution layer*. Thus, the number of parameters of a shared convolution layer is:

$$N_{SC} = k \times k \qquad (4)$$

Assuming that the input has m feature maps, each input feature map is processed by the same filter to produce one feature map. Finally, m output feature maps are produced (same number as input feature maps).

*6) The dilated convolution layer* can obtain larger receptive field by interpolating zero values between the elements of a normal convolution kernel according to the dilated rate, while having the same number of parameters as the normal convolution kernel, as shown on the bottom left of Fig. 5. Assuming that the input shape is b×h×w×m, the kernel size of each filter is k×k, output channel = n, and a bias is used in each filter, then the number of parameters of a normal dilated convolution layer is:

$$N_{Dilated} = m \times k \times k \times n + m \times n \qquad (5)$$

Noting r the dilated rate and k×k the kernel size, the receptive field of dilated convolution layer is ((k-1)×r+1)×((k-1)×r+1)) Although dilated convolutions can capture large scale spatial features, they will lose the information of small objects due to gridding effect.

### C. Proposed Layers

*1) The mixed depth-wise separable convolution layer* (MDSC) is designed to extract and fuse short and long-ranged context information, with only very few parameters. It is a combination of group convolution [60] and depth-wise convolution [56]. As shown in Fig. 4 (b), the MDSC consists of p DSC layers with different kernel sizes. First, the m input feature maps are processed by each DSC, producing p×m feature maps. Then, p×m×n filters with kernel size 1×1 are applied to the p×m feature maps to produce n output feature maps. The total number of parameters of a MDSC layer in CD-FM3SF is:

$$N_{MDSC} = p \times m \times n + \sum_{i=1}^{p} m \times k_i \times k_i \qquad (6)$$

In CD-FM3SF, we set p = 2, $k_1$ = 3 and $k_2$ = 5. The number of parameters of one MDSC is about 34/n (n≥64) times that of a normal convolutional layer.

*2) The shared and dilated convolution residual block layer* (SDRB) is inspired by [61] and redesigned to avoid the gridding effect caused by dilated convolutions and extract multi-scale spatial features, also called short and long-range context information in computer vision field [61]. First, shared convolution is used to obtain relatively small scale spatial features with only a few parameters (Fig. 5, right-hand side), followed by a dilated convolution to obtain large scale spatial features. The combination of a shared convolution and a dilated convolution forms a basic layer in SDRB block. Assuming the input shape is b×h×w×m, the kernel size of dilated convolution is $k_i \times k_i$ and a bias is used in each filter, the kernel size of shared conv is $K_i \times K_i$, noting n the number of output channels and $r_i$ the dilated rate, the number of parameters of a SDRB layer is:

$$N_{SDRB} = 2 \sum_{i=1}^{p} (K_i \times K_i + m \times k_i \times k_i \times n + m \times n) \qquad (7)$$

$$K_i = (k_i - 1) \times r_i + 1 \qquad (8)$$

CD-FM3SF includes 6 SDRB blocks (Fig. 3), each with 2 basic layers. The kernel sizes of dilated convolutions in SDRB blocks are set to 3×3, and the rates are set to 2, 2, 3, 3, 4 and 4, from bottom to top. The kernel sizes of shared convolutions are set to 5×5, 5×5 7×7, 7×7, 9×9 and 9×9, respectively. The slow growth rate can help avoid gridding effect caused by dilated convolution [61]. The number of parameters of one SDRB is about $9/K_i^2$ ($K_i \geq 5$) times that of a normal residual block with the same receptive field.

### D. Proposed Concatenation and Sum (CS) Operation

In the proposed CD-FM3SF, the residual architecture is adopted and improved to fuse and pass as much spectral and spatial information as possible. In the encoder, each branch includes a convolution layer and a mixed depth-wise separable convolution layer (MDSC) followed by a max-pooling layer (Fig. 3). In the top branch, the inputs and outputs of MDSC are summed at pixel level. In the lower two branches, the output of the convolution layer is first concatenated with the output of the max-pooling layer of the upper branch at channel level, then the concatenated feature maps are put into MDSC. Because both Conv and Max-pooling layers output 64 feature maps in middle and bottom branches, the number of output feature maps is 128 after concatenating them at channel level. The normal residual operation adding input and output feature maps at pixel level cannot be done. In order to solve this problem, we propose a Concatenation and Sum (CS) operation used in the last two branches, which is also a novelty element of CD-FM3SF.

As mentioned in subsection 3.1, the convolution kernel sizes in MDSC are set to 3 and 5 to extract multi-scale spatial features from the inputs then fuse them together. The concatenation operation can combine the spectral features from previous and current branches (multi-scale spectral features) by concatenating the input feature maps at channel level. The multi-scale spatial features extracted from multi-spectral bands can be fully fused by the MDSC layer following the concatenation operation. In order to introduce the residual operation and transfer the multi-scale spatial and spectral features in the middle and bottom branches as in the top branch, after being processed by MDSC layer, the two inputs of concatenation operation and the output of MDSC are summed at pixel level. This CS operation also can fuse the spectral features from previous and current branches and their fused multi-scale spectral and spatial features again, further improving the usage of the multi-scale spatial and spectral information and passing it onto the next layer.

### E. Details of CD-FM3SF

CNNs have been proved to better preserve spatial and spectral information than human designed interpolation algorithms [62], [63]. Therefore, instead of rescaling all Sentinel-2 bands to the same resolution by explicit interpolation algorithms, we used CNN to process different resolution bands in CD-FM3SF automatically. The proposed CD-FM3SF method is a fully convolution neural network which consists of an encoder (Fig. 3, left-hand side, down-sample part) and a decoder (right-hand side, up-sample part). The encoder consists of three input branches that process the corresponding multi-spectral bands, and the decoder has three output branches that predict three cloud masks with corresponding resolutions of the input bands. The max-pooling layer is used to down-sample the feature maps in the encoder and the Deconv layer is use to up-sample the feature maps in the decoder. The strides of Deconv layer are set the same as that of max-pooling layer to restore the features to the same scale. The number of parameters of a normal convolutional layer is m×k×k×n, with k set to 3 in most deep learning-based models, m and n the number of input and output channels of the convolutional layer, which are usually set to 64, 128, 256 and 512. Thus, the number of parameters of a normal convolutional layer depends more on m and n than k. For Sentinel-2 images, 13 inputs bands feed the input branches of CD-FM3SF and 3 different resolutions (10m, 20m and 60m) are produced in output branch. Therefore, a normal convolutional layer is added as the first layer in each input branch and the last layer in each output branch, that can extract more features from the inputs but with fewer parameters.

Although the structure of the encoder and decoder are similar, there are some differences in details. In the encoder, MDSC and SDRB are used to obtain multi-scale spatial features, while in the decoder, DSC is adopted to fuse features from the same level in encoder and decoder. Because large scale spatial features have already been extracted in the encoder, there is no need to increase the kernel size in the decoder to obtain large scale spatial features. Therefore, the decoder only needs to decode the features extracted by the encoder, which is also the reason why CS operation is only used in the encoder.

The skip connection of U-Net is adopted in CD-FM3SF. The output feature maps of the encoder are concatenated with that of the decoder at channel level, which enables the model to fuse the features from the encoder and decoder during the up-sampling of each level. The fusion of low-level features can help the network restore more high-resolution details lost during down-sampling, thus improving image segmentation accuracy. Skip connections can also help CD-FM3SF converge faster [64]. The basic layers in SDRB, DSC and MDSC are followed by batch norm [65] and Rectified Linear Unit (ReLU) activation function.

The cross-entropy loss function [66] has proved very effective in image segmentation task [67], [68], thus was used as the loss function of CD-FM3SF (eq. 9):

$$L = -\frac{1}{w \times h} \sum_{i=1}^{w} \sum_{j=1}^{h} \left( y_{i,j} \ln \bar{y}_{i,j} + (1 - y_{i,j}) \ln(1 - \bar{y}_{i,j}) \right) \qquad (9)$$

where y is the reference cloud mask, $\bar{y}$ is the predicted mask. w and h are the width and height of the reference mask.

In order to better detect clouds with different sizes, CD-FM3SF is designed to produce three cloud masks at different spatial resolutions of 10 m, 20 m and 60 m respectively. The total loss is defined as follows:

$$L_{total} = L(y_t, \bar{y}_t) + 0.1L(y_m, \bar{y}_m) + 0.01L(y_b, \bar{y}_b) \qquad (10)$$

where t, m and b indices refer to top, middle and bottom branches, respectively. $y_t$ can help CD-FM3SF detect small clouds and produce accurate cloud edges. With the supervision of $y_m$, the previous layers before the output layer of $y_m$ can extract more useful information of middle size clouds then pass this information to the upper layers. Similarly, $y_b$ can help previous layers (before its output layer) extract more useful information of large clouds and pass it onto the upper layers. Because CD-FM3SF aims to produce accurate cloud masks at the highest resolution available from inputs, the importance of output cloud masks is related to their resolutions. Therefore, the weights of losses are set to 1, 0.1 and 0.01 for top, middle and bottom branches, respectively.

$y_m$ and $y_b$ are used to improve the accuracy of $y_t$ on middle and large clouds detection in training stage, thus, in testing stage, we only compare $\bar{y}_t$ with the reference mask at highest resolution available (10m) to evaluate the performance of CD-FM3SF qualitatively and quantitatively.

### F. Baseline Methods

Two traditional cloud detection methods and two deep learning-based methods were compared with the proposed CD-FM3SF. Fmask [9], [10] is a rule-based cloud detection method which has been adopted as the official cloud detection algorithm by the United States Geological Survey (USGS). Fmask 4.0 [12] was developed for cloud detection in Landsat 4-8 and Sentinel-2 imagery. Sen2Cor [21], [69] is the official cloud masking method for Sentinel-2 images. We compared the latest Fmask 4.0 and Sen2Cor 2.8.0 versions with the proposed CD-FM3SF method. Both Fmask 4.0 and Sen2Cor 2.8.0 methods were used with default parameters and in the remainder of this article, those versions and parameter sets are referred to as Fmask and Sen2Cor, respectively.

MSCFF [38] is a deep learning-based cloud detection method which was designed for high and medium resolution images. MSCFF is also a lightweight model that uses similar components as the proposed CD-FM3SF, such as dilated convolution to detect large clouds and a residual architecture to boost network convergence during training. RS-Net [37] was proposed for cloud detection in Landsat-8 images and based on U-Net [58] architecture, which is also the basic framework of CD-FM3SF. Comparing CD-FM3SF to both MSCFF and RS-Net will allow to evaluate the added-value of the proposed method compared to deep learning methods using similar architecture elements.

## IV. EXPERIMENTAL RESULTS

### A. Experimental setting

*1) Hyper-parameters setting*: The hyper-parameters were set to the same for all deep learning-based methods. The batch size and epoch for training were set to 24 and 40, respectively. Adam-optimizer [70] was adopted to optimize the parameters of all networks, with the following hyper-parameters: $\beta_1 = 0.5$, $\beta_2 = 0.9$, and the initial learning rate = 0.001. Exponential decay with decay rate = 0.995 and decay step = 5 was used as learning rate decay strategy. All deep learning-based methods were trained with the TensorFlow version 1.14.0 running on Linux operating system, on 2 Intel(R) Xeon(R) E5-2640 v4 x86_64, 2.4GHz with 20 cores and 4 Nvidia Tesla V100 with 16 GB memory. It took 5 to 8 days for training each deep learning-based methods.

*2) Accuracy metrics*: In order to evaluate the performance of CD-FM3SF quantitatively, 5 accuracy metrics were selected:

overall accuracy (OA), Precision or producer's accuracy (PA), Recall or user's accuracy (UA), F1-score and intersection over union (IoU). The formulas of accuracy metrics are as follows：

$$OA = \frac{TP + TN}{TP + TN + FP + FN} \tag{11}$$

$$Precision = \frac{TP}{TP + FP} \tag{12}$$

$$Recall = \frac{TP}{TP + FN} \tag{13}$$

$$F1\text{-}score = \frac{2 \times Precision \times Recall}{Precision + Recall} \tag{14}$$

$$IoU = \frac{TP}{TP + FP + FN} \tag{15}$$

where TP, TN, FP and FN are the numbers of correctly predicted cloud pixels, correctly predicted clear pixels, wrongly predicted cloud pixels and wrongly predicted clear pixels, respectively. OA is an indicator used to evaluate the percentage of correctly predicted pixels with regards to the total pixels, precision and recall are related to commission and omission errors, F1-score is a comprehensive index that combines precision and recall, and IoU is the ratio of the intersection and union of the true and predicted cloud regions. OA, F1-score and IoU are comprehensive indicators that can evaluate the classification performance more accurately than Precision and Recall alone.

In addition, to better evaluate the robustness of CD-FM3SF and baseline methods quantitatively, we used the UA-PA curve [71] and the TPR-FPR or Receiver Operating Characteristic (ROC) curve [72], with true positive rate (TPR) and false positive rate (FPR) calculated as follows:

$$TPR = \frac{TP}{TP + FN} \tag{16}$$

$$FPR = \frac{FP}{FP + TN} \tag{17}$$

User accuracy (UA), producer accuracy (PA), TPR and FPR were computed for the cloud detection results of all methods were calculated with thresholds ranging from 0 to 1, respectively, then the area under each UA-PA and ROC curves were computed. The bigger the areas under UA-PA and ROC curves, the better the robustness of the method. Because Fmask and Sen2Cor only produce binary masks, they are represented by a single point on these curves.

### B. Comparison with Baseline Methods

*1) Quantitative and qualitative results*: The test experiments were conducted on 8 Sentinel-2A images from the WHUS2-CD dataset, evenly distributed over China mainland and covering about 96000 km² earth surface (Table II). The accuracy metrics were applied to all cloud detection results of CD-FM3SF and baseline methods on all test images, and quantitative results are summarized in Table III. It can be seen that MSCFF-13 and RS-Net-13 achieve higher values on most accuracy metrics than MSCFF and RS-Net, respectively. However, CD-FM3SF still obtains the best performance on OA, Precision, F1-score and

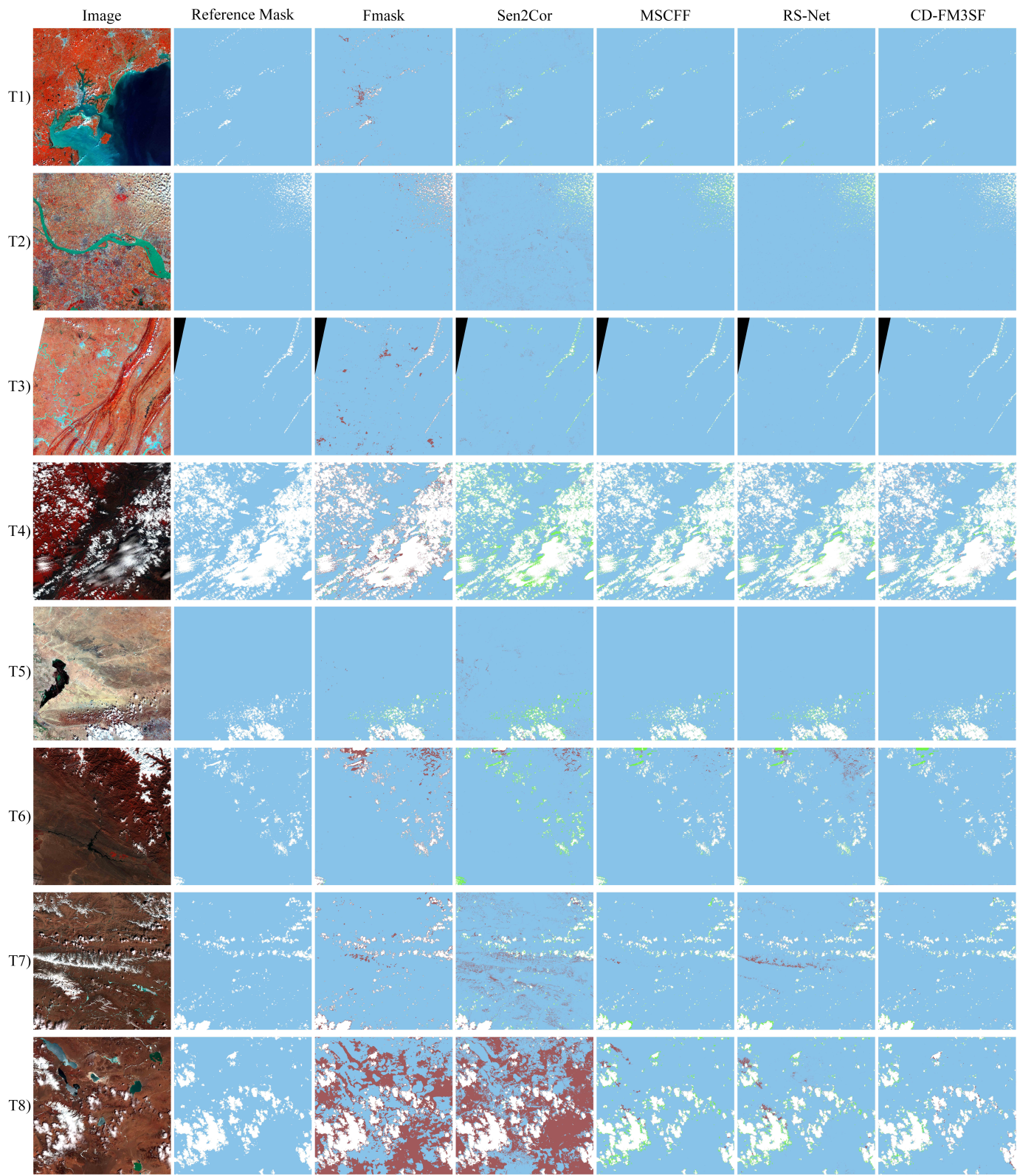

Fig. 6. Cloud detection results of different methods for all testing Sentinel-2A images. T1…T8 refers to the whole testing images in Table II. Commission errors are in red, omission errors in green. Pixels in agreement with reference are in blue for background and white for clouds, and black is no data. The first column shows the composited images with bands 8/3/2.

Fmask. The IoU value demonstrates that CD-FM3SF can achieve better balance between commission and omission errors than baseline methods. In order to better illustrate this, we enhanced the cloud detection results of all methods visually and present them in Fig. 6. Light blue pixels represent correctly predicted background, white pixels correctly predicted clouds, dark red pixels committed clouds (e.g. background wrongly classified as clouds) and green pixels omitted clouds (i.e. clouds wrongly classified as background). The less dark red and green pixels in the visually enhanced results, the better cloud detection results.

TABLE III
AVERAGE ACCURACY INDICATORS OF DIFFERENT METHODS ON TESTING IMAGES T1…T8. THE HIGHEST VALUES ARE MARKED IN BOLD.

| Methods | OA | Precision | Recall | F1-score | IoU |
|---|---|---|---|---|---|
| Fmask | 92.62% | 59.90% | **95.14%** | 0.7098 | 0.5713 |
| Sen2Cor | 90.89% | 64.20% | 69.36% | 0.6252 | 0.4698 |
| MSCFF | 98.40% | 96.42% | 78.77% | 0.8637 | 0.7645 |
| MSCFF-13 | 98.78% | 96.22% | 86.48% | 0.9102 | 0.8363 |
| RS-Net | 98.03% | 88.77% | 79.90% | 0.8372 | 0.7259 |
| RS-Net-13 | 98.59% | 93.63% | 85.06% | 0.8900 | 0.8038 |
| CD-FM3SF | **98.86%** | **96.50%** | 87.75% | **0.9186** | **0.8503** |

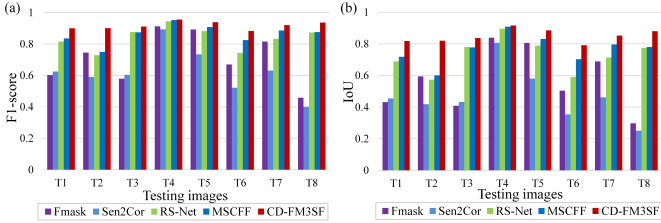

Fig. 7. Distributions of F1-score (a) and IoU (b) by WHUS2-CD testing images and different methods.

Fig. 6 shows the results of cloud masking methods on all testing images T1 to T8 of WHUS2-CD dataset, at full scale. Fmask always produces more cloud labels than other methods, leading to commission errors (dark red pixels) and decreased precision as shown in Table III. Cloud detection results of CD-FM3SF show less commission errors and omission errors (green pixels) than baseline methods, and are therefore in better agreement with reference cloud masks. This is consistent with CD-FM3SF having the highest IoU value among all methods (Table III). Sen2Cor had the most commission and omission errors of all compared methods. Both Fmask and Sen2Cor did not perform well on highlights such as building in T1/T2, snow in T6/T7 and barren land in T8. It should be noted that dark red pixels almost fill the whole cloud detection results of Fmask and Sen2Cor in T8, showing poor performance on barren land. MSCFF and RS-Net fail to distinguish snow from clouds in T6/T7/T8 images, unlike CD-FM3SF in these testing images.

In order to quantitatively evaluate the methods performances over different land cover types, cloud detection results were compared to manually delineated cloud masks in each testing image of the WHUS2-CD dataset, and the F1-score and IoU value were calculated. The results in Fig. 7 demonstrate that deep learning-based methods always achieved better performance than Fmask and Sen2Cor for F1-score and IoU. In each testing image, CD-FM3SF obtains the highest F1-score among all methods (Fig. 7 (a)). In addition, CD-FM3SF obtains the most significant added-value compared to other deep learning-based methods on T1/T2/T6/T8 (water/barren/snow). CD-FM3SF obtains a higher IoU than baseline methods over all testing images (Fig. 7 (b)), with a significant performance gap for all testing images except T4. Fig. 8 shows cloud detection results of all methods for some local regions from WHUS2-CD dataset, over different land cover types whose reflectance is lower than that of clouds in VNIR bands. We can observe that for all methods, there are only few commission errors the cloud detection results, i.e. these land cover types can be easily distinguished from clouds by all methods. The results of Fmask always include commission errors (or false alarms), i.e. Fmask commits more clouds than those in the reference masks. In contrast, Sen2Cor always include more omission errors than other methods. It should be noted that the cloud detection results of MSCFF and RS-Net in water regions (Fig. 8 e)) contain significantly more omission errors than Fmask, Sen2Cor and CD-FM3SF.

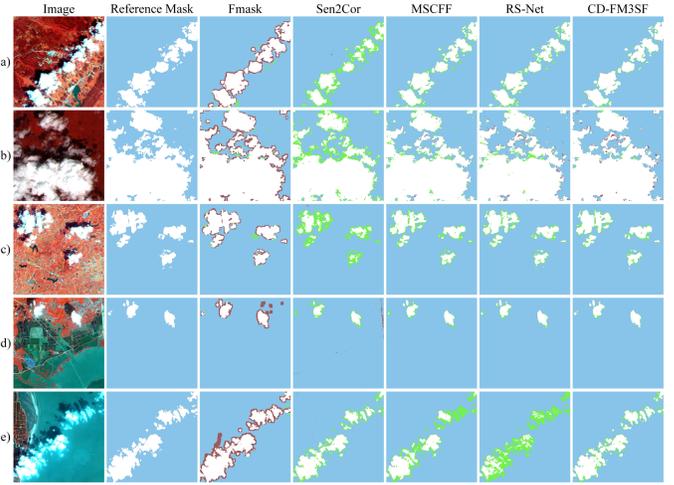

Fig. 8. Examples of cloud detection results for various land cover types with reflectance lower than clouds in VNIR bands. (a) forest. (b) grass. (c) shrubland. (d) wetland. (e) water. The first column shows composited images with bands 8/3/2. Commission errors are in red, omission errors in green. Pixels in agreement with reference are in blue for background and white for clouds.

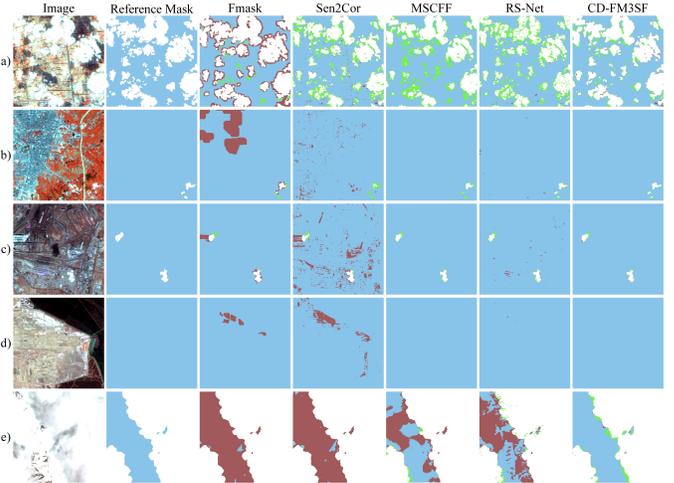

Fig. 9. Examples of cloud detection results for various land cover types with high reflectance similar to that of clouds. (a) farmland in the autumn. (b) buildings with vegetation. (c) buildings with barren. (d) barren. (e) snow/ice. The first column shows composited images with bands 8/3/2. Commission errors are in red, omission errors in green. Pixels in agreement with reference are in blue for background and white for clouds.

Fig. 9 shows cloud detection results of all methods for some local regions in WHUS2-CD dataset with different land cover types whose reflectance is similar to that of clouds. There were more commission errors in cloud detection results of Fmask and Sen2Cor than those of deep learning-based methods. It was more difficult for Fmask and Sen2Cor to distinguish highlights such as buildings and ice/snow from clouds, despite e.g. recent development included in Fmask 4.0 to reduce commission errors over these bright land surfaces [12]. RS-Net also classified some highlights as clouds in those same land cover types. CD-FM3SF and MSCFF perform almost similar on these land cover types, with a lower number of commission errors. However, MSCFF produced slightly more omission errors than CD-FM3SF, which indicates that CD-FM3SF can detect more clouds than MSCFF.

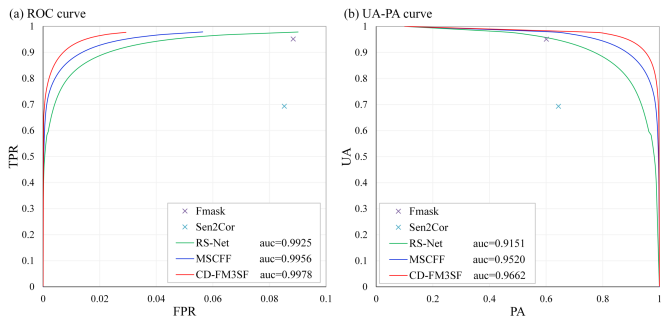

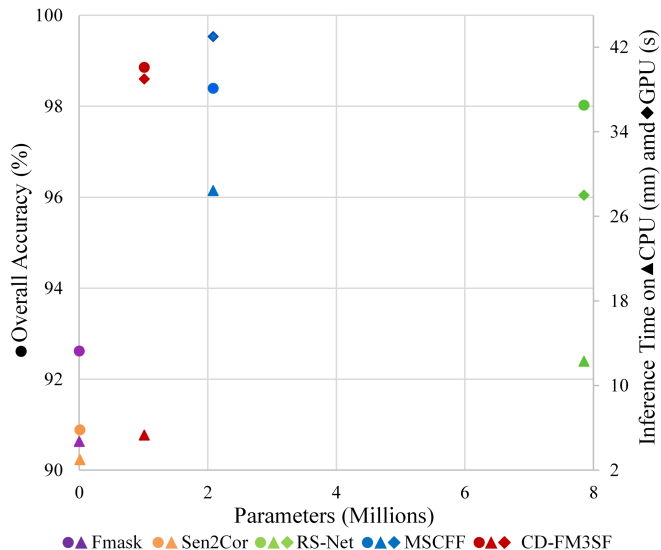

Fig. 10. ROC curves (a) and UA-PA curves (b) of compared methods. Fmask and Sen2Cor are represented by a single cross point.

In general, traditional methods using spectral information to detect clouds such as Fmask and Sen2Cor perform well in vegetation/water areas with lower reflectance than clouds in VNIR bands (Fig. 8). However, Fmask and Sen2Cor often failed in land cover types whose reflectance can be as high as clouds, e.g. snow or buildings (Fig. 9). Deep learning-based methods like MSCFF and RS-Net are very effective in most land cover types such as building and barren, however, they cannot always distinguish snow from clouds with only VNIR bands as inputs. CD-FM3SF which takes SWIR and atmospheric vapour bands into consideration not only performs the best on building and barren but also on snow. Although SWIR and atmospheric vapour bands have lower resolution in Sentinel-2A images, they can really be beneficial to distinguish snow from clouds in VNIR bands. An ablation experiment is proposed in Section 4.5 to quantify the added-value of SWIR and atmospheric vapour bands in CD-FM3SF for cloud detection, and allow a more fair comparison with baseline deep learning methods with only VNIR bands as inputs.

Fig. 10 shows ROC curves and UA-PA curves for the compared methods. The points representing Fmask and Sen2Cor are under the curves of all deep learning-based methods, showing lower overall performance. Compared to MSCFF and RS-Net, CD-FM3SF obtains a slightly higher area under curve (auc = 0.9978) for the ROC curve and a significantly higher auc = 0.9662 for the UA-PA curve, which indicates higher robustness of the proposed method. The TPR value of CD-FM3SF is saturated when FPR value is 0.03, while the saturated values of the TPR values of MSCFF and RS-Net are 0.055 and 0.09, respectively (Fig. 10 b)). The UA value of CD-FM3SF start to decrease for PA value higher than 0.8, while the UA values of MSCFF and RS-Net start to decrease for PA values higher than 0.6 and 0.1, respectively. This indicates that the cloud probability values of CD-FM3SF are more concentrated around 0 and 1 than those of MSCFF and

RS-Net. In general, CD-FM3SF is more robust than MSCFF and RS-Net for cloud detection in WHUS2-CD dataset.

*2) Model complexity and efficiency*: The average computing time for the 8 Sentinel-2A testing images of WHUS2-CD dataset was compared to evaluate the efficiency of the different methods. The deep learning-based methods were tested on GPU node with a single Nvidia Tesla V100, 16 GB memory. Since Fmask and Sen2Cor are not optimized for GPU computing, they were tested on an AMD Ryzen 9 3950X 16-Core CPU processor. It has been shown that depth-wise separable convolution layer (DSC) is not more efficient than normal convolution layer on GPU, even with much less parameters [73]. Therefore, we also tested deep learning-based methods on CPU for computing time. The computing time of CD-FM3SF and compared methods is shown in Fig. 11, which also includes the number of parameters of deep learning-based methods.

Deep learning-based methods cost much less computing time on GPU than Fmask and Sen2Cor at inference, and more on CPU. Overall Fmask costs the longest computing time (284 s), about ten times longer than RS-Net on GPU. CD-FM3SF costs the second shortest time on GPU (39 s), between RS-Net (28 s) and MSCFF (43 s). Although CD-FM3SF is only the second fastest on GPU, it is the fastest deep model on CPU, twice faster than RS-Net, and achieves the best overall accuracy. Furthermore, while the original U-Net model has 28 million parameters, CD-FM3SF only has 1.01 million parameters, which is only about 3.6% of the number of U-Net parameters and less than half the size of the lightweight model MSCFF, which makes it a very light deep learning model. The faster computing times of RS-Net compared to the lighter CD-FM3SF model on GPU can be partly explained by native optimizations in Tensorflow for standard layers used in RS-Net, whereas the newly designed layers and operations in CD-FM3SF (MDSC layer, SDRB layer and CS operation) could not be equally well optimized in Tensorflow yet. In summary, CD-FM3SF has less

Fig. 11. Model complexity and efficiency. The horizontal axis is the number of parameters in the methods, the left vertical axis is overall accuracy in %, the right vertical axis are inference time on CPU (in minutes) and on GPU (in seconds).

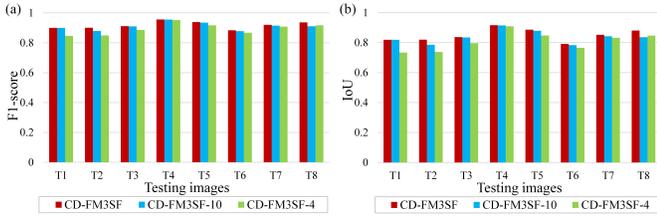

Fig. 12. Distributions of F1-score (a) and IoU (b) by WHUS2-CD testing images for CD-FM3SF based methods.

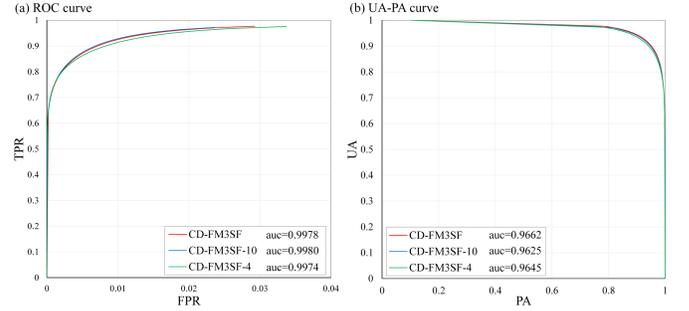

Fig. 13. The ROC and UA-PA curves of CD-FM3SF-based methods. (a) ROC curves. (b) UA-PA curves.

TABLE IV

AVERAGE ACCURACY INDICATORS OF CD-FM3SF BASED METHODS ON ALL TESTING IMAGES. THE HIGHEST VALUES ARE MARKED IN BOLD.

| Methods | OA | Precision | Recall | F1-score | IoU |
|---|---|---|---|---|---|
| CD-FM3SF | **98.86%** | 96.50% | **87.75%** | **0.9186** | **0.8503** |
| CD-FM3SF-10 | 98.74% | 97.14% | 85.76% | 0.9105 | 0.8366 |
| CD-FM3SF-4 | 98.68% | **97.96%** | 82.44% | 0.8940 | 0.8100 |

than half the number of parameters, costs less than half the inference time on CPU and achieves higher overall accuracy than other deep learning-based methods.

### C. Influence of Input Spectral Bands

Ablation experiments were conducted on the inputs, to analyze the influence of input spectral bands on CD-FM3SF for cloud detection and allow a fair comparison with baseline deep learning methods that use only VNIR bands. CD-FM3SF-10 is the version of CD-FM3SF without atmospheric vapour bands (Ca/WV/Cir) as inputs and CD-FM3SF-4 is the version of CD-FM3SF with only VNIR bands as inputs. Table IV shows the quantitative cloud detection results of CD-FM3SF based methods, averaged over the 8 WHUS2-CD testing images. The performance of CD-FM3SF based methods increases as the input spectral information increases. CD-FM3SF performs the best on OA, Recall, F1-score and IoU values. However, CD-FM3SF-4 achieves 97.96% precision value which is slightly higher than that of CD-FM3SF. This is because without the supervision of large scale clouds, CD-FM3SF-4 will detect less clouds around the large cloud boundary. Combining Table III and IV, it can be seen that CD-FM3SF-4 performs better than MSCFF and RS-Net on OA, Precision, Recall, F1-score and IoU values. This indicates that CD-FM3SF based methods are more effective than MSCFF and RS-Net for cloud detection in Sentinel-2A images even with only VNIR bands as inputs.

CD-FM3SF always obtains slightly higher values for F1-score and IoU than CD-FM3SF-10 and CD-FM3SF-4 over all testing images (Fig. 12). This is because with more useful spectral information as input, CD-FM3SF can detect clouds more accurately than CD-FM3SF-10 and CD- FM3SF-4. In most testing images, the advantage of CD-FM3SF over CD-FM3SF-10 and CD-FM3SF-4 is higher for IoU than F1-score.

The ROC and UA-PA curves of all CD-FM3SF based methods are similar (Fig. 13), with very close auc. The ROC and UA-PA curves of CD-FM3SF-4 are slightly lower than those of CD-FM3SF-10 and CD-FM3SF. Combining Fig. 10 and 14, we can observe that all CD-FM3SF based methods have higher auc values for ROC and UA-PA curves than MSCFF and RS-Net. This also demonstrates that the cloud probability values produced by CD-FM3SF based methods are more concentrated around 0 or 1 than MSCFF and RS-Net. Therefore, the CD-FM3SF based methods are more robust than MSCFF and RS-Net, even when the same spectral bands are used as inputs in all methods.

Fig. 14 shows results of all compared methods over different land cover types. Results of methods based on CD-FM3SF are very similar in most cases, except for snow/ice (Fig. 14 i)). For vegetation and water areas (Fig. 14 a) to f)), commission errors (dark red pixels) and omission errors (green) in cloud detection results of CD-FM3SF based methods are fewer than those of Fmask and Sen2Cor. Fmask, Sen2Cor, MSCFF and RS-Net misclassify building as clouds (Fig. 14 g)), while CD-FM3SF based methods do not commit building as clouds. Most clouds in barren samples are omitted by Fmask, Sen2Cor, MSCFF and RS-Net (Fig. 14 h), however, CD-FM3SF based methods detect more clouds than baseline methods. Both CD-FM3SF-10 and CD-FM3SF-4 can distinguish clouds from snow/ice correctly. However, there are more commission errors (dark red pixels) in snow/ice area in CD-FM3SF-4 results than those of CD-FM3SF-10 and CD-FM3SF, which indicates that without atmospheric vapour and SWIR bands as inputs, CD-FM3SF-4 cannot distinguish clouds from snow/ice very well. Although CD-FM3SF-4 performs worse than CD-FM3SF-10 and CD-FM3SF in snow/ice, it misclassifies less snow/ice as clouds than Fmask, Sen2Cor, MSCFF and RS-Net.

### D. Comparison on Snow/Ice Scenes

From Figs. 6, 8, 9, 13 and 14, we can see that distinguishing clouds from highlights such as snow/ice is much more difficult than distinguishing clouds from other low reflectance land cover types such as vegetation. Therefore in this section, we emphasize quantitative results for snow/ice scenes. The averaged accuracy metrics of T6/T7/T8 and 4 additional snow/ice scenes (Fig. 15) are shown in Table V.

CD-FM3SF obtains the highest OA, Precision, F1-score and IoU values among all methods on snow/ice scenes, while Fmask performs the worst on these scenes. CD-FM3SF-4 performs better than MSCFF-13 and RS-Net-13 on all accuracy metrics except Recall. With SWIR as inputs, CD-FM3SF is more effective than CD-FM3SF-4 on these snow/ice scenes. The wide performance gap between CD-FM3SF and baseline methods on the most comprehensive accuracy measures F1-score and IoU shows the added value and robustness of the proposed method on challenging snow/ice scenes.

We also calculated quantitative results on 12 testing images

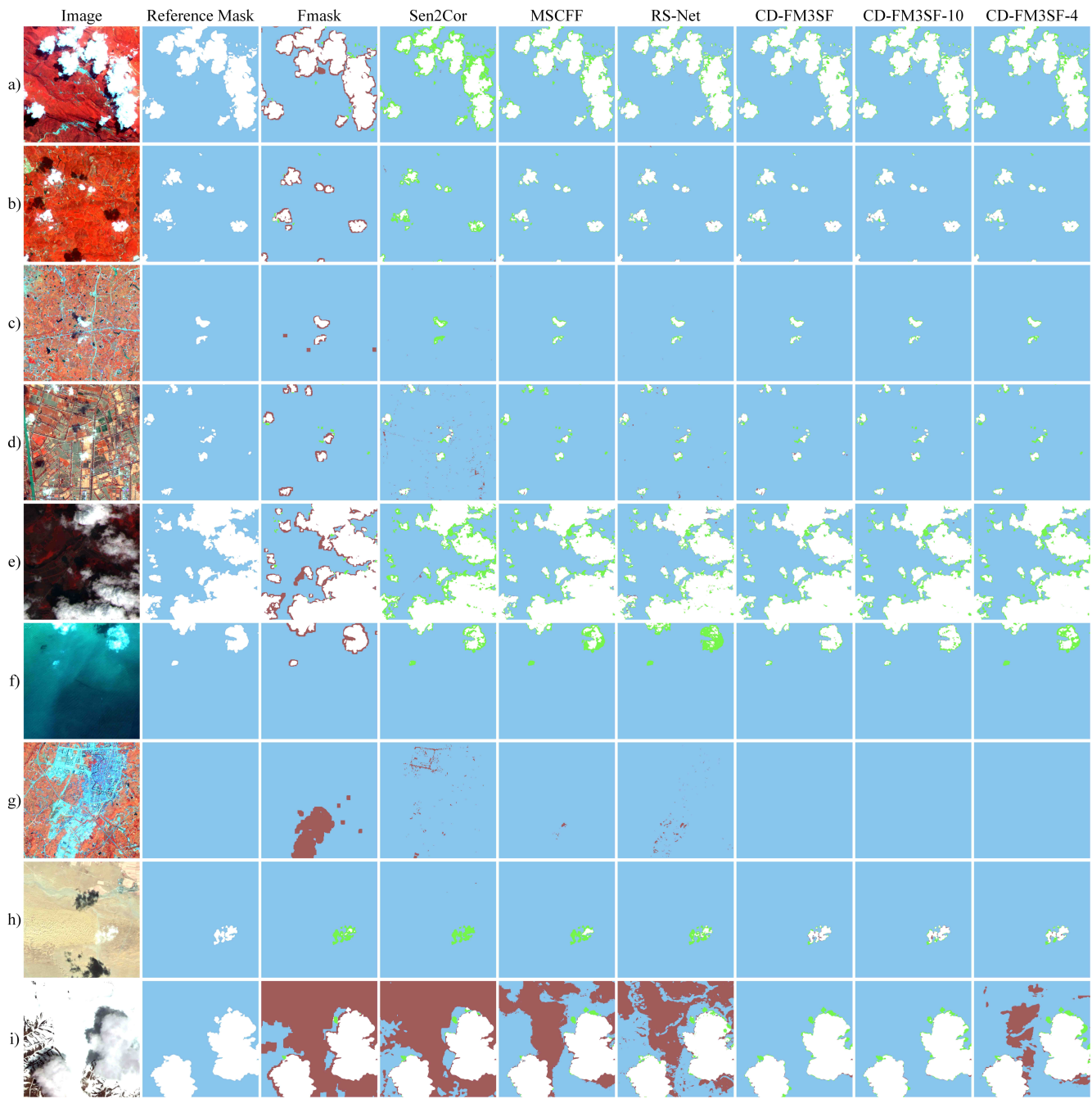

Fig. 14. Examples of cloud detection results for different land cover types by CD-FM3SF-based methods. (a) forest. (b) grass. (c) shrubland. (d) farmland. (e) wetland. (f) water. (g) building. (h) barren. (i) snow/ice. The first column shows the composited images with bands 8/3/2. Commission errors are in red, omission errors in green. Pixels in agreement with reference are in blue for background and white for clouds.

TABLE V

AVERAGE ACCURACY INDICATORS OF DIFFERENT METHODS ON 7 SNOW/ICE TESTING IMAGES IN WHUS2-CD+. THE HIGHEST VALUES ARE MARKED IN BOLD.

| Methods | OA | Precision | Recall | F1-score | IoU |
|---|---|---|---|---|---|
| Fmask | 57.74% | 13.60% | **99.37%** | 0.2393 | 0.1359 |
| Sen2Cor | 72.46% | 18.58% | 92.13% | 0.3092 | 0.1829 |
| MSCFF | 87.66% | 30.84% | 68.00% | 0.4243 | 0.2693 |
| MSCFF-13 | 93.49% | 50.77% | 90.37% | 0.6501 | 0.4816 |
| RS-Net | 77.03% | 18.10% | 68.98% | 0.2867 | 0.1673 |
| RS-Net-13 | 94.11% | 53.70% | 86.47% | 0.6626 | 0.4954 |
| CD-FM3SF | **98.32%** | **89.17%** | 85.23% | **0.8715** | **0.7723** |
| CD-FM3SF-10 | 97.58% | 83.60% | 79.40% | 0.8145 | 0.6870 |
| CD-FM3SF-4 | 97.50% | 88.90% | 71.51% | 0.7927 | 0.6565 |

TABLE VI

AVERAGE ACCURACY INDICATORS OF DIFFERENT METHODS ON 12 TESTING IMAGES IN WHUS2-CD+. THE HIGHEST VALUES ARE MARKED IN BOLD.

| Methods | OA | Precision | Recall | F1-score | IoU |
|---|---|---|---|---|---|
| Fmask | 74.18% | 24.22% | **97.80%** | 0.3883 | 0.2409 |
| Sen2Cor | 82.69% | 30.71% | 84.83% | 0.4509 | 0.2911 |
| MSCFF | 92.27% | 52.55% | 79.63% | 0.6331 | 0.4632 |
| MSCFF-13 | 95.75% | 68.30% | 92.01% | 0.7840 | 0.6447 |
| RS-Net | 85.96% | 35.13% | 79.72% | 0.4877 | 0.3224 |
| RS-Net-13 | 96.07% | 71.05% | 89.64% | 0.7927 | 0.6566 |
| CD-FM3SF | **98.57%** | 93.03% | 89.70% | **0.9132** | **0.8403** |
| CD-FM3SF-10 | 98.14% | 90.91% | 86.43% | 0.8861 | 0.7955 |
| CD-FM3SF-4 | 98.04% | **94.20%** | 81.67% | 0.8749 | 0.7776 |

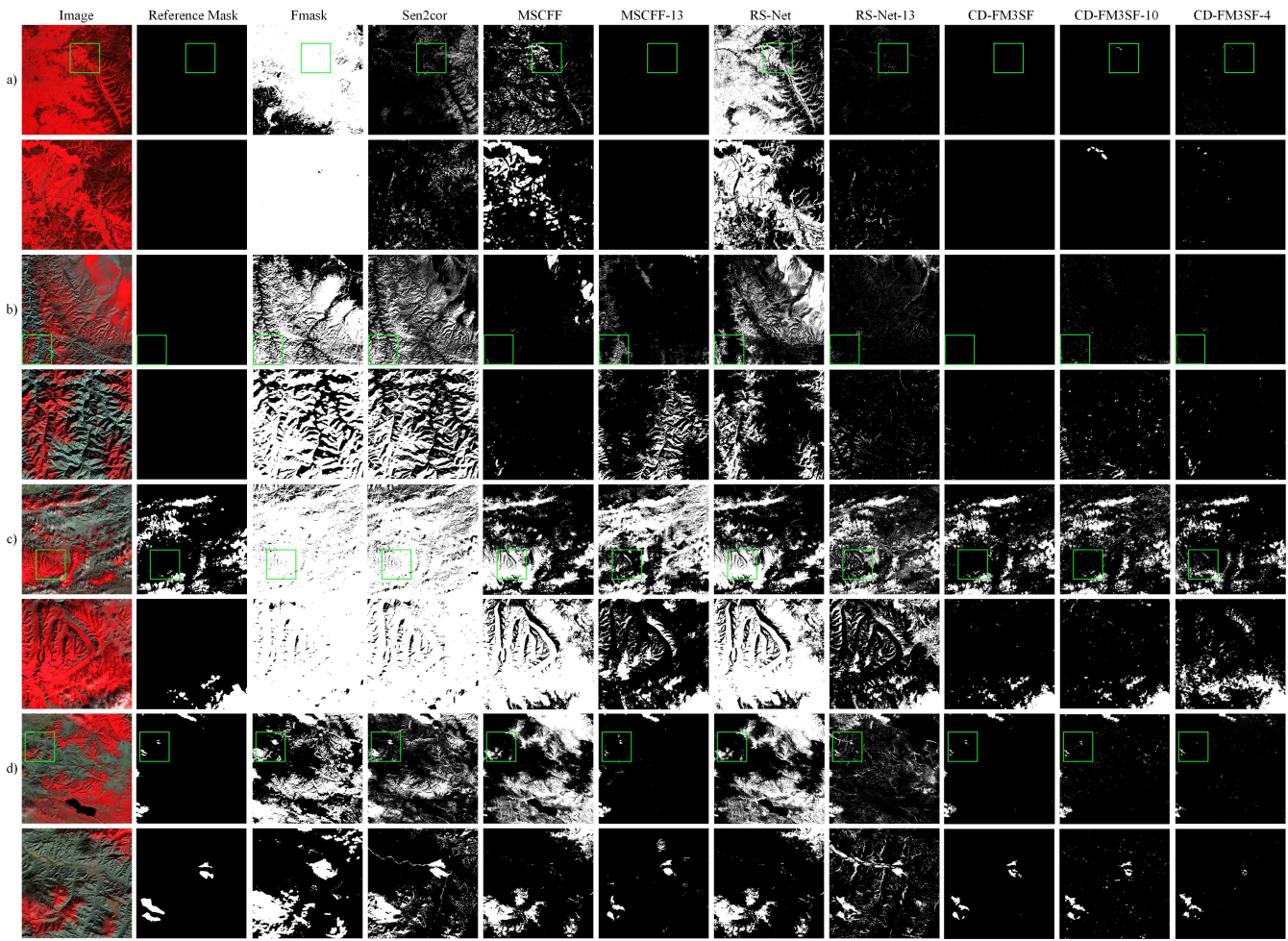

Fig. 15. Cloud detection results with different methods for four additional snow/ice scenes (full scenes on odds rows, zoom-in on green square on even rows) (a) Snow/ice/Mountain/Water, no clouds (S2A_MSIL1C_20210207T023851_N0209_R089_T52UCU). (b): Snow/ice/Mountain/Barren, no clouds (S2A_MSIL1C_20210102T054231_N0209_R005_T43SFB) (c) Snow/ice/Mountain/Barren, clouds(S2A_MSIL1C_20210126T052111_N0209_R062_T44SNE). (d) Snow/ice/Mountain/Water, clouds (S2A_MSIL1C_20201206T041141_N0209_R047_T47SMV). The first column shows the composited images with bands 8A/11/12. Black and white colors denote background and clouds, respectively.

in WHUS2-CD+. Averaged accuracy metrics in Table VI show that CD-FM3SF still obtains better performance than baseline methods on most accuracy metrics. Combining with Table III, CD-FM3SF based methods are more robust than deep learning-based baseline methods.

Fig. 15 shows the visual results of different methods for 4 additional snow/ice scenes. Fig. 15 a) and b) are two cloud-free scenes, c) and d) are two scenes including clouds. It can be seen that Fmask, Sen2Cor MSCFF and RS-Net cannot always distinguish clouds from snow/ice. Even when including SWIR bands, RS-Net-13 mis-classified snow/ice as clouds in all 4 scenes, while MSCFF-13 failed in scenes Fig. 15 b) and c) and was outperformed even by CD-FM3SF-4 that used only VNIR bands. In the most challenging scene Fig. 15 c), all methods mis-classified snow/ice on the mountains as clouds, except CD-FM3SF and CD-FM3SF-10. CD-FM3SF had overall the best ability to distinguish clouds from snow/ice on these 4 scenes, with very few confusion between snow/ice and clouds, especially on the most challenging scenes with mountainous areas covered in snow/ice.

## V. Discussion

### A. Overall Assessment of CD-FM3SF

Generally speaking, the traditional rule-based method Fmask often detected more clouds than other methods in WHUS2-CD, because it combines of multiple empirical decision trees. The threshold-based Sen2Cor method omits more clouds than other methods because the spectral reflectance thresholds used in Sen2Cor are set high. However, Sen2Cor misclassifies more highlights as clouds than Fmask, even though it has high spectral reflectance thresholds. Deep learning-based methods MSCFF and RS-Net performed better on WHUS2-CD dataset than Fmask and Sen2Cor. This is because in addition to spectral information, spatial information used in deep learning-based methods which can help distinguish clouds from highlights when they have similar spectral reflectances. Spectral and spatial information in VNIR bands can be used for accurate cloud detection in most scenes except on snow/ice, because both spectral and spatial information of snow/ice are similar to that of clouds in VNIR bands. All CD-FM3SF based methods achieve better performance than Fmask, Sen2Cor, and original deep learning-based baseline methods MSCFF and RS-Net on WHUS2-CD, which indicates that the architecture of CD-FM3SF is effective and robust for cloud detection in Sentinel-2A images, even with limited spectral information from VNIR bands. The experiment results of CD-FM3SF based methods indicate that the spectral information in atmospheric vapour and SWIR bands can help distinguish clouds from snow/ice even they have low spatial resolution.

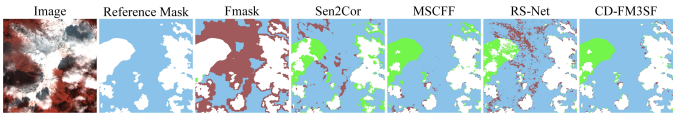

| Image | Reference Mask | Fmask | Sen2Cor | MSCFF | RS-Net | CD-FM3SF |

Fig. 16. Cloud detection results of different methods for a very thin cloud sample in WHUS2-CD (T6 test image). Commission errors are in red, omission errors in green. Pixels in agreement with reference are in blue for background and white for clouds.

When including all 13 spectral bands as inputs to the baseline deep learning methods through interpolation (MSCFF-13 and RS-Net-13), average accuracy measures over all 8 test images in WHUS2-CD improved (Table III). And MSCFF-13 performed better on average than CD-FM3SF-4 and CD-FM3SF-10, and only marginally worse than CD-FM3SF. However, the quantitative results on 7 snow/ice scenes in WHUS2-CD+ show that CD-FM3SF has a much better ability than MSCFF-13 to distinguish clouds from snow/ice.

A detailed visual comparison on 4 additional challenging scenes containing snow/ice and mountainous areas (Fig. 15) clearly showed that CD-FM3SF-based methods were more robust and consistent over these conditions than all other methods, with much less confusion between clouds and snow/ice over mountains. In particular the proposed CD-FM3SF lightweight cloud detection method clearly outperformed the best lightweight model in our baseline (MSCFF-13) on these challenging scenes, which demonstrates the added value of the proposed method.

### B. Limitation and Future Perspectives

Although the proposed CD-FM3SF method achieves better performance than baseline cloud detection methods for various land cover types, CD-FM3SF shares a limitation with most deep learning-based methods. Because there are only a few very thin cloud samples in WHUS2-CD dataset, deep learning-based methods fail to detect very thin clouds. As shown in Fig. 16, very thin clouds are omitted by Sen2Cor, MSCFF, RS-Net and CD-FM3SF. Although Fmask mis-classifies snow/ice as clouds, it detects very thin clouds accurately.

Labelling more very thin clouds would be a way to solve this problem. However, the boundaries of this kind of clouds are challenging to label for humans. In our previous work on cloud removal, we combined generative adversarial network with a modified physical model (CR-GAN-PM, [24]) to use unpaired clear and thin cloud images to separate background and clouds. Cloud layers produced by CR-GAN-PM can be treated as cloud probability maps of thin clouds. In future work, we will combine CD-FM3SF with CR-GAN-PM to detect very thin clouds without the need to label thin clouds at pixel level.

In this work, CD-FM3SF was designed to detect clouds. Since clouds generate shadows, cloud shadow detection is also necessary in remote sensing images. Manual labelling of cloud shadows brings additional challenges, e.g. spectral similarity or mixing with several land cover classes. That is one of the reasons why most Sentinel-2 cloud detection datasets do not include cloud shadow labels, especially not at the highest resolution available (10m). A possible approach would be to manually refine shadow masks produced by either Fmask or Sen2Cor, although this could introduce biases towards either method in testing stage. Another potential solution could be to use approximate labels in training, by performing erosion on shadow masks produced by either Fmask or Sen2Cor. This weak supervision strategy, that increases the likelihood of correctly labelled reference pixels, was independently proposed by [74] for VHR semantic segmentation and [75] for cloud and shadow detection in Landsat images.

Since convolutional layers can re-sample images to specific resolution by setting suitable stride, CD-FM3SF can be modified to support other imaging satellites with bands at different resolutions such as CBERS-04, ZY-1 02D and HJ-1B, as well as the widely used Landsat and GF-2 satellites.

## VI. CONCLUSION

In this paper, we proposed a lightweight cloud detection method fusing multi-scale spectral and spatial features (CD-FM3SF) in Sentinel-2A multi-spectral images. CD-FM3SF aims to make full use of the spectral information in all Sentinel-2A bands and improve cloud detection accuracy.

In order to evaluate the performance of CD-FM3SF on cloud detection, a new annotated Sentinel-2 cloud detection dataset over China was created, WHUS2-CD, with 32 images covering various land cover types and all seasons, and reference cloud masks delineated manually at the highest resolution available (10 m). We further added 4 snow/ice images to WHUS2-CD (WHUS2-CD+) to allow more accurate validation of cloud masking methods, and comprehensive testing of their robustness under diverse and challenging conditions. The WHUS2-CD+ cloud detection dataset has been released as open data in order to contribute to an easier comparison of cloud masking methods on Sentinel-2 images.

The experiment results on WHUS2-CD+ dataset show that CD-FM3SF is much more effective and robust than traditional methods, especially in distinguishing clouds from clouds. Compared to deep learning-based methods, CD-FM3SF achieves much higher accuracy with much fewer parameters. The ablation experiment results demonstrate that taking into consideration more spectral bands even with less spatial information can improve the performance of CD-FM3SF on cloud detection. CD-FM3SF can also outperform baseline deep learning-based methods with only VNIR bands as input, which suggests CD-FM3SF could perform well in other satellite images including VNIR bands. Results of inference time on CPU show that CD-FM3SF is more suitable for practical applications than other deep learning-based methods, because most devices are not equipped with a powerful GPU.

In future work, we will extend CD-FM3SF to multi-classes classification such as cloud, cloud shadow, water and snow. A combination of generative adversarial network (GAN) and CD-FM3SF will be taken into consideration to reduce the requirement of labelling cloud masks, which is very time-consuming to obtain especially for thin clouds and cloud shadows. Extension to other satellites will also be considered.

## ACKNOWLEDGMENT

The authors are grateful for the Sentinel-2 data services from the Copernicus Open Access Hub. We would like to thank J.Q. Zhang, Y.S. Wang, G.T. Shen, S.B. Zheng, S.J. Huang, H.Y. Lin, J.H. Luo and R.X. Fang for labelling the WHUS2-CD dataset. The numerical calculations in this paper have been done on the supercomputing system in the Supercomputing Center of Wuhan University.